\documentclass[sigconf]{acmart}
\AtBeginDocument{%
  }

\copyrightyear{2026}
\acmYear{2026}
\setcopyright{cc}
\setcctype{by}
\acmConference[ICSE-FoSE]{2026 IEEE/ACM 48th International Conference on Software Engineering: Future of Software Engineering }{April 12--18, 2026}{Rio de Janeiro, Brazil}
\acmBooktitle{2026 IEEE/ACM 48th International Conference on Software Engineering: Future of Software Engineering (ICSE-FoSE), April 12--18, 2026, Rio de Janeiro, Brazil}
\acmPrice{}
\acmDOI{10.1145/3793657.3793904}
\acmISBN{979-8-4007-2479-4/2026/04}

\begin{document}

\title{Maintaining the Heterogeneity in the \\Organization of Software Engineering Research}


\author{Yang Yue}
\affiliation{%
  \institution{California State University, San Marcos}
  \city{San Marcos, CA}
  \country{USA}}
\email{yyue@csusm.edu}

\author{Zheng Jiang and Yi Wang}
\affiliation{%
  \institution{Beijing University of Posts and Telecommunications}
  \city{Beijing}
  \country{China}}
\email{{jiangzheng|yiwang}@bupt.edu.cn}

\renewcommand{\shortauthors}{Y. Yue, Z. Jiang, and Y. Wang}

\begin{abstract}
  The heterogeneity in the organization of software engineering (SE) research historically exists, i.e., funded research model and hands-on model, which makes software engineering become a thriving interdisciplinary field in the last 50 years. However, the funded research model is becoming dominant in SE research recently, indicating such heterogeneity has been seriously and systematically threatened. In this essay, we first explain why the heterogeneity is needed in the organization of SE research, then present the current trend of SE research nowadays, as well as the consequences and potential futures. The choice is at our hands, and we urge our community to seriously consider maintaining the heterogeneity in the organization of software engineering research.
\end{abstract}

\ccsdesc[500]{Software and its engineering}
\ccsdesc[300]{General and reference~Empirical studies}
\ccsdesc[500]{Social and professional topics~Funding}
\keywords{Organization of Research, Funded Research Model, Hands-on Research Model}
\begin{teaserfigure}
  \includegraphics[width=\textwidth]{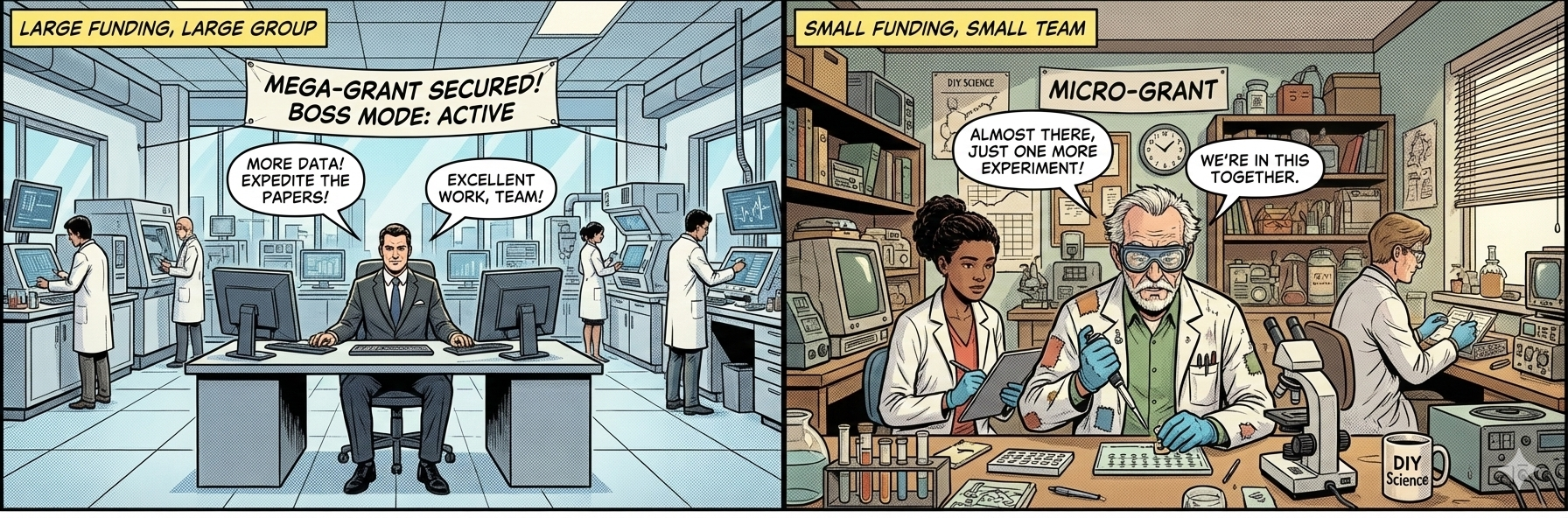}
  \caption{A comic of two research models, funded research model vs. hands-on research model. Picture was generated with Google's Nano Banana Pro).}
  \label{fig:teaser}
\end{teaserfigure}


\maketitle

\section{Introduction}
Think about two software engineering researchers at leading research universities. The first researcher (the left in Fig. \ref{fig:teaser}) runs a 1,000+ square meters lab, hiring over 50 people (PhD students, PostDocs, etc.), and brings in millions of dollars of funding each year. He or she may have published over 800 papers with 100+ coauthors, most of which have long author lists, earning over 40,000 citations and an H-index of 110, and holding a number of patents. The second researcher (the right in Fig. \ref{fig:teaser}) typically works with a handful of collaborators and PhD students, only seeks research funding when necessary. He or she has published about 100 articles, mostly with one or two coauthors or even solo, with about 8,000 citations and an H-index of 40. 

You may think it is purely a personal choice, or, let's be more honest with ourselves, we may assume that the second is less capable than the first. A majority of readers of this essay may believe the first researcher is better than the second. But they are not necessarily true. Such significant differences are basically the two fundamental forms of the organization of research work \cite{Rahmandad19}. The first is the funded research model, where the researchers act as bosses, obtaining funding to sustain a sizable research team. They often do not participate in the actual research but rather oversee those who do. The second is a hands-on research model, where researchers are extensively involved in executing research, train a few graduate students and postdocs, work in small teams, and demand not that much funding. Usually, such heterogeneity happens among different fields. For example, the funded research model is dominant in fields such as biology and engineering, while the hands-on research model is prevalent in mathematics, physics, social sciences, humanities, and management.

There are many reasons why research work takes these two different forms, e.g., the demand for individual talents, the cost of training a qualified research worker, the scale of the engineering work, and so on \cite{tripodi2025tenure}. In this essay, we have no intention to discuss these reasons. However, software engineering was born with both forms of organizations of research. The term ``engineering'' indicates that the funded research model is suitable for building (large) software systems. Meanwhile, we must admit that it also inherited the hands-on research model of research in mathematics; for example, formal methods were basically an application domain of mathematics and logic. In addition, management and social sciences also partially shape the landscape of software engineering, for example, project management and team collaboration. Particularly, in their classic book ``\emph{Peopleware: Productive Projects and Teams},'' Timothy Lister and Tom DeMarco \cite{demarco2013peopleware} argued: \emph{The major problems of our work are not so much technological as sociological in nature}.

Such a mixture of academic traditions contributed to the heterogeneity in the organization of software engineering research, making it a thriving and vibrant interdisciplinary area over the last 50 years. However, we observe that such heterogeneity has been seriously and systematically threatened recently. Failures in acknowledging and maintaining such heterogeneity, in our opinion, at least contribute to a number of problems in our community, e.g., ill-trained students, bad academic tastes, unhealthy publication culture, and gaps with industrial practices.

We thus argue the necessity of maintaining such heterogeneity in this essay. We are going to first explain why software engineering needs heterogeneity in the organization of research work, and how the heterogeneity has been systematically threatened in current software engineering research. We then imagine a few possible futures with(out) such heterogeneity.

\section{Why Software Engineering Research Needs Heterogeneity?}

\subsection{The Nature of Software Engineering Discipline Requires Heterogeneity}
Fred Brooks, the author of The Mythical Man-Month and SE's only Turing Award  laureate, pointed out that:

\begin{quote}
    \textit{``There is no single development, in either technology or management technique, which by itself promises even one order-of-magnitude improvement in productivity, reliability, or simplicity.''} \cite{Brooks87}
\end{quote}

Although he did not explicitly emphasize heterogeneity in software engineering, his work provides a strong foundation for it. Due to the complexity, conformity, changeability, and invisibility of software products, it is impossible to have universal solutions in software engineering \cite{Brooks87}, which is what makes SE different from traditional engineering disciplines such as civil or mechanical engineering, where almost every piece of detail can be precisely computed before the execution based on Physical laws. For example, once the design of a mechanical engine is finalized, it can be produced precisely according to its specification. But for software, uncertainty and unpredictability permeate virtually every aspect of software development \cite{garlan2010software}. Therefore, heterogeneity is required by the nature of the software engineering discipline.

In the organization of the research work in software engineering, heterogeneity is also prevalent. In general, the funded research model and the hands-on research model are two typical models in the research institutes \cite{Rahmandad19}. With a funded research model, principal investigators (PIs) tend to play the role of manager and entrepreneur, i.e., securing funding to support the research team and overseeing the research process. Research teams in the funded research model tend to be large. However, PIs in hands-on research model are completely different; they usually engage with the research process directly, supervise smaller research teams, and have fewer funding demands. 

\subsection{Large Teams Develop, but Small Teams Disrupt}

Although large research teams with a funded research model are dominant in the engineering field \cite{Rahmandad19}, small research teams of the hands-on research model still exist, and both are essential in knowledge discovery through research. Researchers investigated the different sizes of research teams and the character of their research findings, and revealed that smaller teams tend to disrupt science and technology with new ideas and opportunities, while larger teams tend to develop existing ones \cite{wu_large_2019}. This finding is also applicable in software engineering research. One example is Roy Fielding's work of Representational State Transfer (REST), which was developed mostly by a two-person team: Roy and his advisor Dick Taylor, is a profound reconceptualization of large-scale distributed system architecture. The core abstraction of REST emerged from a smaller research team, as ``disruptive work'' of the time, and later was refined, standardized, and scaled by larger teams across academia and industrial institutes. Meanwhile, while large teams often build upon recent, popular ideas, small teams go deeper with less popular ideas. Although large teams may have greater immediate impact, e.g., more papers and citations, small teams' outputs are often more novel and have a durable impact. 

\section{A Reluctant Reality or A Tragedy}

\subsection{The Reality of SE Research: Funding, Paper, Citations, and All Quantifiable Things}

As studies showed, the funded research model is dominant in engineering, which is the reality and the current trend of SE research. First, funding is overly emphasized. The SE research tends to be increasingly organized around externally funded projects. Lots of PIs have to spend a substantial amount of time securing funding to support their research, rather than focusing on research. In addition, funding is also often the most important factor, if not the only one, in faculty evaluation; thus, many junior faculty members have to align their research topics with funding agencies' priorities to receive funding support, rather than their research interests or expertise. Second, the funding and evaluation mechanisms tend to favor incremental, publishable results, since such achievements/contributions are easier to quantify. For example, the amount of received funding and the number of published papers are simple metrics to compare across various research teams and PIs, and the powerful people on Capitol Hill like numbers. As a result, researchers tend to prioritize research projects with quick-turnaround outputs, and the research questions are often scoped to fit the deadlines of SE conferences. This creates implicit pressure towards developmental rather than disruptive work. Third, SE conferences and journals have become more selective and competitive. Every year, SE conferences such as ICSE, FSE, and ASE receive increasing numbers of submissions; however, the number of accepted papers does not increase at the same pace. Given the funding and publication pressures faced by the researchers, it is not surprising to observe more conservative topic selection and increasing homogeneity in methods and problems studied. 

In the survey, some participants also shared similar concerns when answering the question ``Q9: What would you change and what outcome from that change would you like to see in the software engineering research community?''  

\begin{quote}
    ``\textit{One thing? The effort/cost and difficulty behind (Lessons learn, time costs, more research planning)... Bolder research questions, not 'I adapted my RQ to my result'.}'' (No. 98 response to Q9)

    ``\textit{Reduce focus on publication quantity and up quality. Raise the bar to publish thoughtful insightful papers that might impact how software is actually engineered in the future.}'' (No. 103 response to Q9)

    ``\textit{Slow down the pace. The expectations for students are ridiculous. Submitting less but higher quality works will help the reviewing process, as there will be more time to give more thorough reviews and feedback, further improve paper quality and avoid unnecessary re-submissions of rejected papers. This needs to be systematic. Universities should develop incentives to reward quality over quantity.}'' (No. 186 response to Q9)
\end{quote}   
as well as ``Q11: What aspects of being in the software engineering research community cause the greatest amount of stress for you?''
\begin{quote}
    ``\textit{Failed funding bids [especially] large ones with industry I led! [cause the greatest stress]}'' (No. 132 response to Q11)
\end{quote}

This trend would eventually reinforce a feedback loop, i.e., funded work to published work to more funding, and systematically put the hands-on research model and smaller research teams into disadvantages.

\subsection{Consequences}

With the trend of the SE research community being increasingly dominated by the funded research model, the heterogeneity in the SE research community would be significantly reduced. We can anticipate some consequences, or some consequences are already happening. First, junior researcher (e.g., doctoral students and postdocs) training is shifting from apprenticeship to project staffing. PIs tend to spend less time mentoring junior researchers, and of course, they cannot afford to do this; instead, they prefer training them as project staff who are able to execute the funding proposal. The junior researchers would have fewer opportunities to learn how to develop independent research ideas, form meaningful research questions, and engage in a long-term hands-on research process, which would make them less prepared for exploratory research. Second, SE researchers would prefer ``safe'' methods in their research; this would reduce methodological diversity. This would limit researchers' ability to identify and address new problems and challenges in software engineering.

\subsection{The Extinction of Action Research}

Such a tragedy is actually happening. Shifting to the funded research model has been kicking out many essential parts of software engineering. An example is the extinction of action research in software engineering. 
A few decades ago, action research was a popular research method in software engineering. In Easterbrook et al.'s book chapter ``\emph{Selecting Empirical Methods for Software Engineering Research}'' published in 2008 \cite{Easterbrook2008}, action research is introduced as one of five major empirical methods in software engineering. Such a method follows a post-positivist research philosophy grounded in critical thinking. It is driven by real-world problems, and develops useful solutions through participatory research in an iterative fashion \cite{10.1145/2647648.2647656}. But how long has it been since you last saw an action research paper appear in the research track of three major SE conferences, i.e., ASE, FSE, and ICSE? Maybe five or ten years ago? The reason is simple: action research requires a lot of hands-on activities and a long research cycle. It is not cost-effective for the funded research model. Moreover, it is also much easier to fail. A research team may not be able to develop useful solutions for a real-world problem after working with industrial partners for two years. Risk-aversion embedded in the funded research model makes action research no longer a rational choice in selecting empirical methods, at least, mining online data is much safer and more predictable.


\section{Futures at the Crossing or We Have Already Passed It?}

Now, we are at the crossing that leads to two distinct futures. In one future, funded research model will eventually eliminate hands-on research model. In the other, hands-on research model may still have a chance to survive, of course, with the collective efforts from the software engineering community.

\begin{figure}
    \centering
    \includegraphics[width=\linewidth]{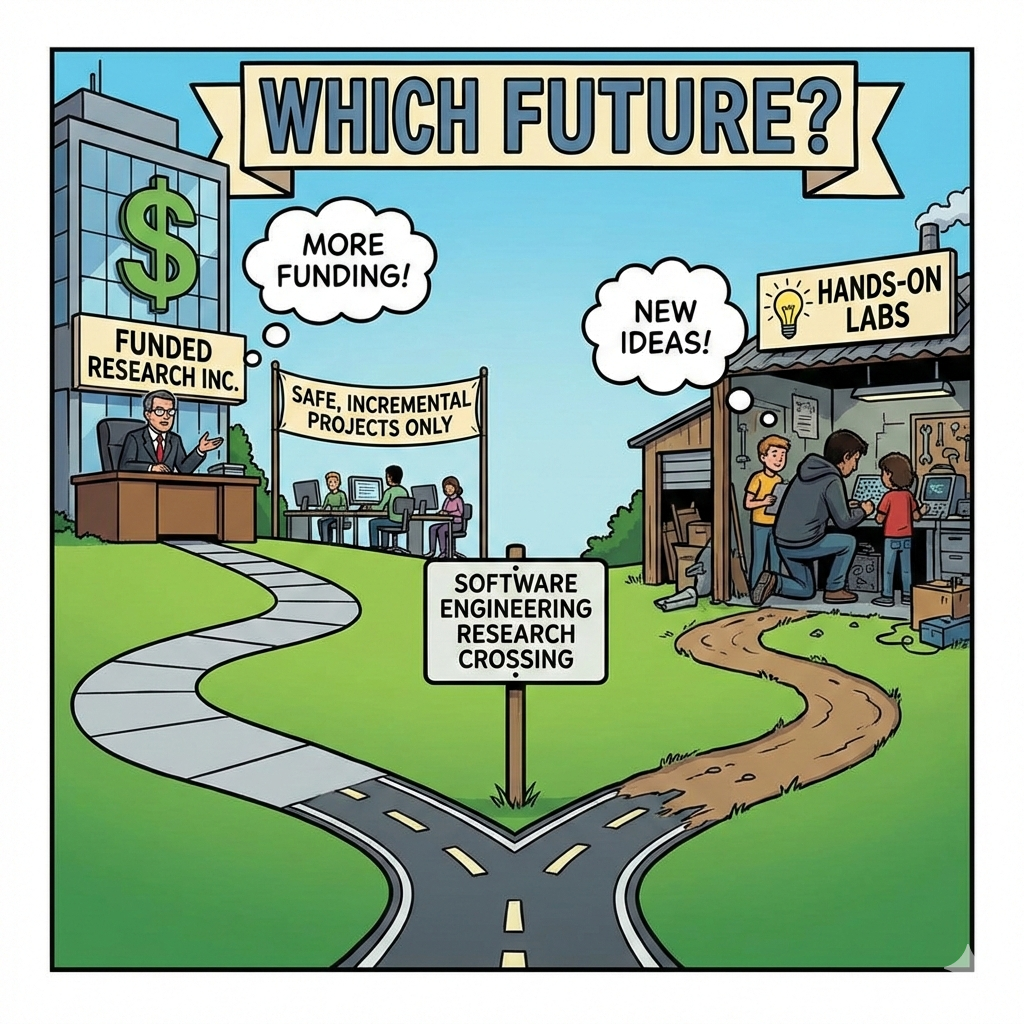}
    \caption{Two possible futures of SE research. Picture was generated with Google's Nano Banana Pro.}
    \label{fig:placeholder}
\end{figure}

\subsection{A Reluctant Future}

In this future, the hands-on research model will no longer exist in the SE research community. While it is excellent for scaling established ideas and making incremental contributions, the dominance of the funded research model could lead to a future where we lose the very spark that drives the field forward. Doing SE research may be similar to production with an industrial assembly line. Such a future may have three features:

\noindent \textbf{\emph{Safe Incrementalism}}. With the disappearance of small, nimble teams that are willing to fail, research shifts toward low-risk bets. Large teams often prioritize ``X+Y'' style papers, for example, LLM + code generation/requirement analysis/program comprehension/fault location..., combining existing techniques to guarantee positive results for stakeholders. This creates a high volume of publications; you may search Google Scholar by yourself, but a scarcity of disruptive innovations that could fundamentally change how we build software. Meanwhile, the long-lasting, fundamental, and \emph{hard} problems in software engineering remain unsolved. For example, research into software development cost estimation has been almost dead for the last 20 years. While the industry has been building very different systems, the methods for software development cost estimation remain in the late 1990s.

\noindent \textbf{\emph{The ``Specialized Cog'' PhD}}. Under large lab scenarios, PhD students may become hyper-specialized cogs in a larger machine. A student might spend four years optimizing a few specific algorithms without ever learning how to frame a research problem, design a user study, or write a full system from scratch. They graduate as expert technicians rather than independent researchers. They may also fail to develop a solid understanding of the fundamentals of software engineering because their advisors may not afford to let them spend a few semesters learning SE fundamentals, which may not be immediately useful in their multi-million dollar project. That happens now! An author of this essay recently received a review for his paper submitted to ICSE, in which the reviewer asked what is ``architecture tactics.'' It is hilarious, right? There must be some failures in the system to make it so amazing, either in how we train researchers or in how ICSE selects PC members.  

\noindent \textbf{\emph{Disconnection with Research and Practices}}. As PIs move further from the real-world research and closer to the spreadsheet (and social media too), the ``reality gap'' widens. We lose the intuitive leap that only comes from a PI getting their hands dirty. The research problems are often something imagined based on others' work. The research becomes something that solves a problem elegantly (often with exhaustive evaluation on standard benchmarks) that never happens \cite{rhodes2025junkification}. In contrast, some real problems are often neglected because the PIs overseeing the research projects have not experienced the friction of a modern production in a decade.

\subsection{An Alternative Future}

In an alternative future, the hands-on research model may survive. It is not totally impossible, and may happen in an unexpected way. Why? The current shrinkage of research funding in major economies and the Matthew effect in funding allocations may eventually exclude a large proportion of researchers from receiving substantial funding, which in turn forces them to choose the hands-on research model. Meanwhile, researchers, particularly those who are more conscientious and reflective, would eventually realize the importance of maintaining diverse models of research.

\subsection{Now It Is on Your Hands}
Now let's return to the two researchers at the beginning of this essay. At some time in your career, you have to answer: \emph{Who do you want to be? The first or the second?} Most readers would choose to be the second? It is perfectly fine, but we want you know, you do have another choice.

\section{Disclaimer}

The authors would like to make it clear that they are not against any models of organizing software engineering research. We do not think a specific model is better than the other. Both models produce valuable SE knowledge and are vital for a healthy scientific and technological ecosystem. At least one author's research lab is partially running under the funded research model. The central thesis of this essay is straightforward--we need to maintain the heterogeneity in the organization of SE research, which actually works quite well in the authors' own experience, having both models in the same lab.


\bibliographystyle{ACM-Reference-Format}
\bibliography{sample-base}


\end{document}